\documentclass[11pt]{article}

\newcommand{\be}{\begin{equation}}
\newcommand{\ee}{\end{equation}}
\usepackage{amsmath}
\usepackage{amssymb}
\usepackage{latexsym}
\usepackage{epsfig}
\usepackage{color}
\usepackage[makeroom]{cancel}

\newcommand{\bkt}[1]{\left(#1\right)}

\newcommand{\pd}[2]{\frac{\partial #1}{\partial #2}} 
 
\renewcommand{\pd}{\partial}

\newcommand{\RRR}{{\hbox{\rm R\kern-2.35mm R}}}
\newcommand{\p}{\partial}

\def\ZZZ{{\hbox{ Z\kern-1.6mm Z}}}

\newcommand{\nin}[1] {\underline{\phantom{h}}\hskip-6pt {#1}}

\newcommand{\sectiono}[1]{\section{#1}\setcounter{equation}{0}}

      \def\H{{\cal H}}   
      \def\M{{\cal M}}     
      
  \def\T{{\cal T}}    
    
\def\Ä{\varphi}  \def\¿{\varpi}	\def\Ï{\vartheta}

\makeatletter
\def\munderbar#1{\underline{\sbox\tw@{$#1$}\dp\tw@\z@\box\tw@}}
\makeatother

\begin{document}

\begin{titlepage}
%\rightline{January 2016} 
\rightline{\today} 
\rightline{\tt MIT-CTP-4766} 
\begin{center}
\vskip 2.5cm

{\Large \bf {Three-point Functions in Duality-Invariant }}\\  

 \vskip 0.7cm

{\Large \bf {Higher-Derivative Gravity }}\\

 \vskip 2.0cm
{\large {Usman Naseer and Barton Zwiebach}}
\vskip 0.5cm
{\it {Center for Theoretical Physics}}\\
{\it {Massachusetts Institute of Technology}}\\
{\it {Cambridge, MA 02139, USA}}\\[2.5ex]
unaseer@mit.edu, zwiebach@mit.edu\\   

\vskip 2.5cm
{\bf Abstract}

\end{center}

\vskip 0.5cm

\noindent
\begin{narrower}

\baselineskip15pt

Doubled $\alpha'$-geometry is the simplest higher-derivative gravitational
theory  with exact global duality symmetry.  We use the double metric 
formulation of this theory  to compute on-shell three-point functions
to all orders in $\alpha'$.  
 A simple pattern emerges when
comparing with the analogous bosonic and heterotic three-point functions.
As in these theories, the amplitudes factorize.
The theory has no Gauss-Bonnet term, 
but contains a Riemann-cubed interaction to second order in $\alpha'$.

\end{narrower}

\end{titlepage}

\baselineskip=16pt
\parskip=\medskipamount			% space between paragraphs (TeX)

\setcounter{tocdepth}{2}
\tableofcontents

\baselineskip15pt

\sectiono{Introduction}

It is well known that the discrete duality symmetries of toroidally compactified 
string theories imply continuous duality symmetries of the classical effective field theory 
for the  massless string degrees of freedom~\cite{Veneziano:1991ek,Sen:1991zi,9109038}.  
Double field theory formulates the
 higher-dimensional two-derivative  
 massless effective field theory in a way that the duality 
symmetry can be anticipated before dimensional reduction~\cite{Siegel:1993th,Siegel:1993bj,Hull:2009mi,Hohm:2010jy,Hohm:2010pp}. 
When higher-derivative corrections (or $\alpha'$ corrections) 
are included it becomes much harder
to provide a duality covariant formulation.  It is generally expected that as soon
as higher-derivatives are included, all numbers of them are required for 
exact duality invariance.  

At present, there is only one known example of an effective gravitational  
theory with higher-derivatives and
exact duality invariance:  the ``doubled $\alpha'$ geometry" of  Hohm, Siegel and Zwiebach, henceforth called HSZ theory~\cite{Hohm:2013jaa}.
Two key facts about 
 this theory are relevant to our discussion.  First, its spacetime Lagrangian 
is efficiently written in terms of
a double metric $\M$,  an unconstrained version of the generalized metric $\H$ which encodes
the metric $g$ and the antisymmetric field $b$ in a familiar fashion.   
The Lagrangian is cubic in $\M$ and includes terms with up to six derivatives.  
In $\H$ variables, however,   
 the Lagrangian has terms of all orders in derivatives~\cite{Hohm:2015mka}.  Second, HSZ theory is not the low-energy effective field theory of bosonic strings, nor that of heterotic strings.  It does not contain gauge fields, but  due to the Green-Schwarz
modification of the gauge transformations of the $b$ field,  it contains higher-derivative terms
such as a Chern-Simons modification of the field strength 
$H$ for $b$~\cite{Hohm:2014eba}.   
A gauge principle to accommodate higher-derivative corrections of bosonic and heterotic
strings has been investigated in~\cite{Marques:2015vua}.   

The purpose of this paper is to calculate   the simplest amplitudes in HSZ theory; on-shell
three-point amplitudes for the metric and $b$ field.  While this is a relatively simple matter
in any gravitational theory described in terms of a metric and a $b$ field, it is 
a rather nontrivial computation in a theory formulated in terms of a double metric $\M$.\footnote{The computation in terms of $\H$ variables would not be practical, as even the terms with four 
derivatives have not been explicitly written out. }
This is so because metric and $b$ field fluctuations are encoded nontrivially in $\M$ fluctuations
and because $\M$ also contains unfamiliar auxiliary fields.
These amplitudes, not yet known, will be obtained
using the $\M$ field Lagrangian.  The procedure is instructive:  it requires us to obtain the explicit $\alpha'$ expansion of the Lagrangian and to discuss the elimination of auxiliary fields. 
The three-point amplitudes turn out to be simple, suggesting that higher point functions should be calculable.    
We suspect that world-sheet methods will eventually prove superior for the computation of general
amplitudes.  In fact, reference~\cite{Hohm:2013jaa} discussed how the chiral world-sheet theory appears to be a singular limit of the conventional world sheet theory, and  
the recent elaboration in~\cite{Siegel:2015axg} 
goes further in this direction and discusses 
amplitudes.  Our results provide a  test of world-sheet methods for the simplest
case. There are other works on amplitudes motivated by or making use 
of double field theory~\cite{Hohm:2011dz,Boels:2015daa}.   

In both bosonic string theory and heterotic theory, on-shell three-point amplitudes      
factorize into factors that involve left-handed indices and right-handed indices (see eqn.(\ref{bos-het-fact})).  
We show that in HSZ these amplitudes also factorize (see eqn.(\ref{eq:DFTprediction})).  
The explicit form of the
result has implications for the low-energy effective field theory.
In the bosonic string the terms in the low-energy effective action 
needed to reproduce its three-point amplitudes
include Riemann-squared (or Gauss-Bonnet)~\cite{Gross:1985rr,Zwiebach:1985uq}  
and $HHR$ terms to first order in $\alpha'$, and 
Riemann-cubed to second order in $\alpha'$~\cite{Metsaev:1987zx,Metsaev:1986yb}.  
To reproduce the (gravitational) 
heterotic three-point amplitudes the theory 
has only order $\alpha'$ terms: Gauss-Bonnet, $HHR$ and 
a $b$-odd term $b\,  \Gamma \partial \Gamma$, with $\Gamma$ the Christoffel connection. 
 At order $\alpha'$ HSZ theory contains only the $b$-odd term with twice the coefficient in 
heterotic string, and to  second order in $\alpha'$  the bosonic string Riemann-cubed term with {\em opposite} sign.
Our work shows that to order $(\alpha')^2$,  the following is the gauge invariant
action that reproduces the on-shell cubic amplitudes of HSZ theory:  
\be
\label{CubicCov}   
S \ = \   \int d^Dx  \sqrt{-g}\,  e^{-2\phi} \Bigl( R +  4 (\partial \phi)^2 -\tfrac{1}{12} \widehat H_{ijk} \widehat H^{ijk}    -   \, \tfrac{1}{48}\, \alpha'^2 \,  R_{\mu\nu}{}^{\alpha\beta} 
R_{\alpha\beta}{}^{\rho\sigma}R_{\rho\sigma}{}^{\mu\nu}  \Bigr)\;.  
\ee
 The ${\cal O}(\alpha')$ terms above arise from the kinetic term for the  
three-form curvature~\cite{Hohm:2014xsa}.  We have  
$\widehat H_{ijk}  =   H_{ijk}  +  3\,\alpha'  \Omega_{ijk}  ( \Gamma)$,  where $H_{ijk} = 3\, \partial_{[i} b_{jk]}$ with the Chern Simons term $\Omega$ given by:  
\be
\Omega_{ijk}  ( \Gamma) \ = \   \Gamma_{[i | p | }^{\,q} \partial_j  \Gamma_{k] q}^{\,p} + \tfrac{2}{3} 
   \Gamma_{[i | p | }^{\,q}  
   \Gamma_{[j | r | }^{\,p}  
   \Gamma_{[k ] q | }^{\,r} \;. 
\ee
In the conclusion section we discuss possible calculations that may advance our understanding
of duality-invariant higher-derivative field theories. 

%%%
\sectiono{Bosonic, heterotic, and HSZ three-point amplitudes}  

In this section we motivate and state our main claim:  In HSZ theory,  
 on-shell three-point amplitude
for gravity and $b$ fields exhibits a factorization structure analogous to that of the bosonic and heterotic string.  
For this purpose let us consider these amplitudes.  Let $k_1, k_2,$ and $k_3$ denote the
momenta of the particles.  Since we are dealing with massless states, the
on-shell condition and momentum conservation imply that for all values of $a, b = 1,2,3$ : 
\be
\ k_a \cdot k_b = 0 \,. 
\ee
We also have three polarization tensors $e_{a\,  ij}$ with $a= 1,2,3$.
Symmetric traceless polarizations represent  gravitons, and antisymmetric polarizations 
represent  $b$ fields. 
Dilaton states are encoded by polarizations proportional to the 
Minkowski metric~\cite{Scherk:1974ca}.   
 The polarizations satisfy transversality
\be
k_a ^i  e_{a\,  ij} = 0    
  \,,  \quad k_a ^j  e_{a\,  ij} = 0 \,, \ \     \quad  a \ \hbox{not summed.}  
\ee
To construct the three-point amplitudes one defines the auxiliary three-index tensors
$T$ and $W$: 
\be
\label{TW}
\begin{split}
T^{ijk} (k_1, k_2, k_3) \ \equiv \  & \  \eta^{ij}\,  k_{12}^k \ + \ 
 \eta^{jk}\,   k_{23}^i \, 
\ + \ \eta^{ki} \, k_{31}^j \,  
\,, \\[1.0ex]
W^{ijk} (k_1, k_2, k_3)\ \equiv \  & \   \tfrac{1}{8} \, \alpha' \  k_{23}^i\, k_{31}^j  k_{12}^k  \,   
 \,  ,
\end{split}
\ee
with $k_{ab} = k_a - k_b$. Note the  invariance of $T$ and $W$ under simultaneous cyclic shifts
of the spacetime indices and the $1,2,3$ labels.   For  bosonic and heterotic strings the on-shell amplitudes
for three massless closed string states with polarizations $e_{a \, ij}$ 
are given by (see, for example, eqn.(6.6.19) in \cite{Polchinski:1998rq} and eqn. (12.4.14) in \cite{Polchinski:1998rr}):
\be
\label{bos-het-fact}
\begin{split}
S_{bos} \ = \ & \  \tfrac{i}{2} \kappa\,   (2\pi)^D \delta^{D}(\sum p)   \, e_{1ii'} 
e_{2jj'} e_{3kk'} \ ( T + W)^{ijk} \, ( T + W)^{i'j'k'} \,, \\[1.0ex] 
S_{het} \ = \ & \   \tfrac{i}{2} \kappa\,   (2\pi)^D \delta^{D}(\sum p)   \, e_{1ii'} 
e_{2jj'} e_{3kk'} \ ( T + W)^{ijk} \, T ^{i'j'k'}\,. 
\end{split}
\ee
Note the factorization of the amplitude into a factor that involves the first indices on the
polarization tensors and a factor that involves the second indices on the 
polarization tensors.\footnote{The  
 on-shell conditions satisfied by the momenta imply
that there are no candidates for three-point amplitudes with more than six derivatives.} 
 We claim that in HSZ theory  
the on-shell amplitudes also factorize:  
\be
\label{eq:DFTprediction}
S_{hsz} \ = \  
  \tfrac{i}{2}\kappa\,   (2\pi)^D \delta^{D}(\sum p)   \, e_{1ii'} 
e_{2jj'} e_{3kk'} \ ( T +  \, W)^{ijk} \, ( T -  W)^{i'j'k'}\,. 
\ee

For the bosonic string $( T + W)^{ijk} \, ( T + W)^{i'j'k'}$
is symmetric under the simultaneous exchange of primed and unprimed indices.
As a result,  the amplitude
for any odd number of $b$ fields vanishes.  
Expanding out
\be
S_{bos} \, = \  \tfrac{i}{2} \kappa\,  (2\pi)^D \delta^{D}(\sum p)   \, e_{1ii'} 
e_{2jj'} e_{3kk'}  \Bigl( T^{ijk} T^{i'j'k'}  + [ W^{ijk} T^{i'j'k'}   + W^{i'j'k'}  T^{ijk} ]
 +W^{ijk}  W^{i'j'k'}  \Bigr) \,, 
\ee
making clear the separation into two-, four-, and six-derivative structures, all of which
are separately invariant  under the simultaneous exchange of primed and unprimed indices. 
The four-derivative structure indicates the presence of Riemann-squared or 
Gauss-Bonnet terms~\cite{Gross:1985rr,Zwiebach:1985uq}.
The six-derivative structure implies the presence of Riemann-cubed terms~\cite{Metsaev:1986yb}. 
For the heterotic string we write the amplitude as 
\be
\begin{split}
S_{het} \ = \    \tfrac{i}{2} \kappa\,  (2\pi)^D \delta^{D}(\sum p)   \, e_{1ii'} 
e_{2jj'} e_{3kk'} \ \Bigl(  & \ T^{ijk} T^{i'j'k'} \ + \tfrac{1}{2} \, [ W^{ijk} T^{i'j'k'}   + W^{i'j'k'}  T^{ijk} ] \\
&  +\tfrac{1}{2}  [ W^{ijk} T^{i'j'k'}   - W^{i'j'k'}  T^{ijk} ]
 \Bigr)\,. 
\end{split}
\ee
We have split the four-derivative terms into a first group, symmetric
under the simultaneous exchange of primed and unprimed indices, and a second group,
antisymmetric under the simultaneous exchange of primed and unprimed indices.
The first group is  one-half of the four-derivative terms in bosonic string theory, a well-known
result.
The second group represents four-derivative terms that can only have an odd number of $b$ fields.  
In fact, only one $b$ field is allowed.  The term with three $b$ fields would 
have to be of the form $HH\partial H$, with $H=db$ and it can be shown to vanish by Bianchi identities.   
The term that one gets is of the form $H \Gamma \partial \Gamma$, and arises 
from the kinetic term of the Chern-Simons corrected $b$-field field strength.   This kind of term
also appears in HSZ theory, as  discussed in~\cite{Hohm:2015mka}. 

Expanding the  HSZ amplitude above one finds  
\be
\label{hsz-expanded}
S_{hsz}  = \, 
  \tfrac{i}{2}\kappa\,   (2\pi)^D \delta^{D}(\sum p)   \, e_{1ii'} 
e_{2jj'} e_{3kk'}  \Bigl(  T^{ijk} T^{i'j'k'}  +  [ W^{ijk} T^{i'j'k'}   -  T^{ijk} W^{i'j'k'} ]   - W^{ijk} W^{i'j'k'} \Bigr)\,, 
\ee
implying that there is no Gauss-Bonnet term, that the term with four derivatives 
has a single $b$ field and is the same as in heterotic string but with twice the magnitude.
The six-derivative term is the same as in bosonic string, but with opposite sign.
This implies that the Riemann-cubed term  in the HSZ action and in bosonic strings 
have opposite signs.  Most of the work in the rest of the paper deals with the computation
of the $g$ and $b$ three-point amplitudes that 
confirms (\ref{hsz-expanded}) holds.   

It is useful to have simplified expressions for the amplitudes.  For later use we record the following
results, with `cyc.'  indicating that two copies of the terms to the left must be added with 
cyclic permutations 
 of the 1,2, and 3 labels:
\be
\label{partial-amplitudes}
\begin{split}
 e_{1ii'} e_{2jj'} e_{3kk'} T^{ijk} T^{i'j'k'}  = \ & \  \hbox{tr} (e_1^Te_2) (k_{12} e_3 k_{12}) 
 + k_{12} ( e_3 e_2^T e_1 + e_3^T e_2 e_1^T) k_{23} + \hbox{cyc.}
   \\[0.5ex]
  e_{1ii'} e_{2jj'} e_{3kk'} (W^{ijk} T^{i'j'k'} \pm T^{ijk} W^{i'j'k'} )   = \ & \ 
  \tfrac{1}{8} \alpha' \, \bigl[ k_{12} (e_3 e_1^T \pm e_3^T e_1) k_{23}  ( k_{31} e_2 k_{31}) + \hbox{cyc.} 
  \bigr]\,  ,
  \\[0.5ex]
  e_{1ii'} e_{2jj'} e_{3kk'} W^{ijk} W^{i'j'k'}  = \ & \   \tfrac{1}{64} \alpha'^2 
  ( k_{12} e_3 k_{12})( k_{23} e_1 k_{23}) ( k_{31} e_2 k_{31})\,.  
\end{split}
\ee

The formulae (\ref{bos-het-fact}) for massless 
on-shell three-point amplitudes 
also hold for amplitudes that involve the dilaton. For the dilaton 
one must use a polarization tensor  
proportional to the Minkowski metric.    
Although we will not use the HSZ
action to compute dilaton amplitudes, the predictions from 
the factorized amplitude (\ref{eq:DFTprediction}) 
are exactly what we expect for the  the 
dilaton.  We explain this now.

Let $\hat\phi$ denote the physical dilaton field.    
For  cubic dilaton interactions $\hat\phi^3$ there is no on-shell candidate at two, four, or six derivatives.
For $\hat\phi^2  e$ interactions there is no on-shell candidate at four or six derivatives, 
but there is one at two  derivatives:  $\p^i\hat\phi \p^j \hat\phi \, e_{ij} \sim \p^i\hat\phi \p^j \hat\phi \, h_{ij}$. 
This term does arise from the first line in (\ref{partial-amplitudes})      
when we take 
$e_{1ii'} \sim  \eta_{ii'}\hat\phi$,    
 $e_{2jj'} \sim  \eta_{jj'} \hat \phi$, 
and $e_{3kk'} = h_{kk'}$.   
It is present in all three theories as it is the universal coupling of a scalar to gravity.

For $\hat\phi ee$ there are no on-shell candidates with six derivatives, but there 
are candidates with two and with four derivatives. 
Let's consider first the on-shell candidates with two derivatives.  
Again, an examination of the first line in (\ref{partial-amplitudes})   
shows that $\hat\phi hh$ vanishes.  This is expected:  the physical dilaton  does not
couple to the scalar curvature.  There is also no $\hat\phi h b$ coupling.
On the other hand one can check that $\hat\phi bb$ does not vanish.
This is also expected, as an exponential of $\hat\phi$ multiplies the $b$-field kinetic term.  
Again, all this is valid for the three theories.

Let us now consider $\hat\phi e e$ on-shell couplings with four derivatives. 
There is just one on-shell candidate:   $\hat\phi \p^{ij} e_{kl} \p^{kl} e_{ij}$.  Due to the
commutativity of derivatives this term requires both $e$'s to be gravitons.  This
coupling arises both in bosonic and heterotic string theory
 because an exponential of $\hat\phi$ multiplies Riemann-squared terms.
 As expected, 
 it can be seen from the second line in (\ref{partial-amplitudes}), using the top sign. 
 It does not arise in HSZ theory because in this theory the four-derivative
 terms are odd under the $\mathbb{Z}_2$ transformation $b \to - b$~\cite{Hohm:2014xsa}, 
 and thus must involve a $b$ field. In conclusion, 
 HSZ theory only has on-shell couplings of dilatons
 at two derivatives, and shares them with heterotic and bosonic strings.  The latter
 two have a single on-shell coupling of the dilaton at four derivatives.   These are
 indeed the  predictions of the three factorized formulae.

\sectiono{Derivative expansion of HSZ theory}

\noindent
Our first goal is to give the action for $\M$ and $\phi$ in explicit form and organized
by the number of derivatives, a number that can be zero, two, four, and six.  
While the parts with zero and two derivatives are known and take
relatively simple forms~\cite{Hohm:2013jaa,Hohm:2014xsa},   
 the parts with
four and six derivatives are considerably longer.  We give their partially simplified   
forms and then their fully simplified forms
when the dilaton field is set to zero.  This will suffice for our later computation
of on-shell three-point amplitudes for gravity and $b$ field fluctuations.

\noindent
We will define actions $S$ as integrals over the double coordinates of the density $e^\phi$
times the Lagrangian $L$.  For the theory in question \cite{Hohm:2013jaa} we have
\be
\label{action_lagrangian}
S = \int e^\phi L \,, \qquad  L \ = \ \tfrac{1}{2} \hbox{tr} (\T ) \, - \, \tfrac{1}{6} \, 
\langle \T | \T \star \T \rangle \,.
\ee
The field $\T$ is a tensor operator and encodes the double metric. 
For arbitrary tensor operators $T$ we have the expansion 
\be
T \ =  \   \tfrac{1}{2}  \, T^{MN} Z_M Z_N  - \tfrac{1}{2}  ( \hat T^M Z_M ) ' \,,
\ee
here $T^{MN}$  and $\hat T^M$ are, respectively, the tensor part and  the pseudo-vector part of the tensor operator. 
The trace of the tensor operator $T$ is a scalar operator  tr$\, T$ defined by (eqn.(5.17), \cite{Hohm:2013jaa})    
\be
\hbox{tr} \, T \ \equiv  \  \eta^{MN} T_{MN} - 3 ( 
T^{MN} \p_M \p_N\phi      
+ \partial \cdot \hat T
+ \hat T \cdot \partial \phi ) \,.
\ee
If a tensor operator $T$ is divergenceless, the pseudo-vector part is determined as
a dilaton dependent function $G$ linear in 
 the  tensor component:
\be
\hat T^M   \ = \ G^M ( T_{PQ}) \ = \   G_{1}^M (T) + G_{3}^M (T) \, ,    
\ee
where  
$G_{1}$ and $G_{3}$ have one and three derivatives,
respectively (eqn.(5.37), \cite{Hohm:2013jaa}):   
\be 
\begin{split}
G_{1}^M (T) \ = \   & \ \p_N T^{MN}  + T^{MN} \p_N \phi\\[0.5ex]
G_{3}^M (T) \ = \ & -\tfrac{1}{2} T^{NP} \p_N\p_P \p^M \phi  - \tfrac{1}{2} \p^M 
\bigl( \p_N \p_P T^{NP}  + T^{NP} ( \p_N \p_P \phi + \p_N\phi \p_P \phi)\ + 2 \p_N\ T^{NP}\ \p_P \phi \bigr) \,. 
\end{split}
\ee
We make the following remarks:    
\begin{enumerate}
\item The free index on $G_{3}$ is carried by a derivative. 
\item  $G_{1}(T)$ and $G_{3}(T)$ both vanish if the two indices in $T_{MN}$ are carried
by derivatives,
\item $G_{3}(T)$ vanishes if one index on $T_{MN}$ is carried by a derivative.
\end{enumerate}

The tensor operator $\T$ featuring in the action is parametrized by a double metric
$\M^{MN}$, and the pseudo-vector part $\hat \M^M$ is determined by the condition that
$\T$ is divergenceless: 
\be
\T \ = \ \tfrac{1}{2}  \M^{MN}  Z_M Z_N  - \tfrac{1}{2}  ( \hat \M^N Z_M ) '  \,, \quad 
\hat \M^M  \ = \ G^M ( \M) \,.
\ee
A short calculation gives
\be
\hbox{tr} \T \ = \ \eta^{MN}\M_{MN} - 3 \p_M\p_N \M^{MN} - 6 \M^{MN} \p_M\p_N \phi 
- 6 \p_M \M^{MN} \p_N \phi - 3 \M^{MN} \p_M\phi \p_N\phi \,,
\ee
which contains terms linear, quadratic and cubic in fields, and no more than two derivatives.
We now use the star product $\star$ 
of two tensors, which gives a divergenceless tensor,  to define
\be
W \ \equiv \ 
 \T \star \T \ = \ \tfrac{1}{2} W^{MN} Z_M Z_N  - \tfrac{1}{2} \bigl( \hat W^M  Z_M \bigr)' \,, 
\ee
where the last equality defines the components of $W$.  The definition of the
 star product  (\cite{Hohm:2013jaa}, sect.6.2) implies that  
\be
W^{MN} \ \equiv   \  ( \T \circ_2 \T)^{MN} \,,  \quad 
W^M \ \  \equiv      \   G^M ( W^{PQ}) \,,   
\ee
the second following because $W$ is divergenceless.  
The formula for product $\circ_2$ is given in (6.67) of \cite{Hohm:2013jaa}.\footnote{In
\cite{Hohm:2013jaa} symmetrizations or antisymmetrizations carry no weight, in this paper they
do.}  
The field $W^{MN}$  has   an expansion on derivatives, 
\be
W^{MN} \ = \ W_{0}^{MN}  + W_{2}^{MN}  + W_{4}^{MN}  + W_{6}^{MN} \,, 
\ee
which, using the notation $\p_{M_1 \cdots M_k} \equiv \p_{M_1} \cdots \p_{M_k}$, takes the form
\be
\begin{split}
W_{0}{}_{MN} \ = \ &  \ 2 \M_{MK} \M^K{}_N\,, \\[1.0ex]
W_{2}{}_{MN}  \ = \ & - \tfrac{1}{2}  \p_M  \M^{PQ} \p_N \M_{PQ} 
+ \M^{PQ} \p_{PQ} \M_{MN}  + 4 \p_{(M} \M^{KL} \p_L \M_{N)K} \\[1.0ex]
&   - 2 \p_Q\M_M{}^P \p_P \M_N{}^Q 
+ G_{1}^K (\M) \p_K \M_{MN} + 2 \bigl( \p_{(N} G_{1}^K (\M) - \p^K G_{1 (N} (\M) \bigr) \M_{M) K} \,, 
\\[1.0ex]
W_{4}{}_{MN}  \ = \ &  \  \p_{MP} \M^{LK} \p_{NL} \M_K{}^P 
- 2 \p_{K(M} \M^{PQ} \p_{PQ} \M_{N)}{}^K     \\[1.0ex]
& \  + 2 \bigl( \p_{(M} G_{3}^K (\M) - \p^K G_{3 (M} (\M) \bigr) \M_{N) K}  
- 2 \p_{P(M} G_{1}^K (\M) \p_K \M^P{}_{N)} \\[1.0ex]
& + \p_P \bigl( \p_{(M} G_{1Q} (\M) - \p_Q G_{1(M}(\M)  \bigr) \p_{N)} \M^{PQ}\,,  \\[1.0ex]
W_{6}{}_{MN}  \ = \ &  -\tfrac{1}{4} \p_{MPQ} \M^{KL} \p_{NKL} \M^{PQ} 
+ \p_P \bigl( \p_{(M} G_{3Q} (\M) - \p_Q G_{3(M}(\M) \bigr) \p_{N)} \M^{PQ} \\[1.0ex]
& \,  -\tfrac{1}{2}  \p_{PQ(M} G_{1}^K(\M) \p_{N)K} \M^{PQ}  \,. 
\end{split}   
\ee
We note that 
\begin{enumerate}
\item On $W_{4 MN}$ at least one index is carried by a derivative.
\item  On $W_{6 MN}$ both indices are carried by derivatives.  
\end{enumerate}
We now turn to the pseudo-vector components $\hat W^K$ which, by definition are given by
\be
\hat W^K \ = \ G^K (W_{MN}) \ =  \ G^K_{1} ( W_0+ W_2 + W_4 + W_6) + G^K_{3} ( W_0+ W_2 + W_4 + W_6)\,.    
\ee
 It then follows by the remarks that the only terms in $W^K$ are:
\be
\begin{split}
\hat W^K_{1}  \ = \  & \ G^K_{1} ( W_0) \,, \\
\hat W^K_{3} \ = \ & \  G^K_{1}( W_2)  \ + G^K_{3}( W_0)\,, \\
\hat W^K_{5} \ = \ & \  G^K_{1}( W_4)  \  + G^K_{3}( W_2)\,. 
\end{split}
\ee
These are terms with one, three, and five derivatives.  
Note that on  $G_{1} (W_4)$ 
 the free index is on a derivative because it
is an index on $W_4$ and the other index on $W_4$ must be the non-derivative
one to have a non vanishing contribution.  
Thus the free index in $\hat W_5$ is on a derivative.   

It is now possible to evaluate the full Lagrangian in (\ref{action_lagrangian}).  For the cubic term we need the inner product formula that follows 
from eqn.(6.67) of \cite{Hohm:2013jaa}   
\be\begin{split}
\langle T_1 | T_2 \rangle \ = \ & \  \ \tfrac{1}{2} T_1^{PQ} T_{2 PQ} - \p_P T_1^{KL} \p_L T_{2K}{}^P
+ \tfrac{1}{4} \p_{PQ} T_1^{KL} \p_{KL} T_2^{PQ}  \\[1.0ex]
& \  -\tfrac{3}{2} ( \hat T_1^M \hat T_2^N \eta_{MN} - \p_N \hat T_1^M \p_M \hat T_1^N) 
-\tfrac{3}{2} ( \p_P\hat T_1^K 
 T_{2K}{}^P  
+ \p_P\hat T_2^K  T_{1K}{}^P )     
\\[1.0ex]
& \ + \tfrac{3}{4} \bigl( \p_{PQ} \hat T_1^K \p_K T_2^{PQ}  + \p_{PQ} \hat T_2^K  \p_K T_1^{PQ} \bigr) \,.
\end{split}
\ee
This formula must be used for  $T_1 = \T$ and $T_2 = W$.  
A useful identity, easily derived by integration by parts, reads
\be
\int e^\phi \ f^K G_{(1)K} (T) \ = \ \int e^{\phi} ( - \p^P f^K \, T_{KP} ) \,. 
\ee
Using this identity and the earlier results we find the following terms in the Lagrangian
\be
\label{l0l2l4l6}
\begin{split}
L_{0} \ = \ & \ -\tfrac{1}{6}\M_{M N} \M^{ N P} \M_{P}{}^{M}\ +\tfrac{1}{2}\M_{M}{}^{M} \,, \\[1.0ex]
L_{2} \ = \ & \ \  \tfrac{1}{2}\bigl(\M^2-1\bigr)^{MP} 
\M_{P}{}{}^{N} \p_{M}\p_{N}\phi\ +\tfrac{1}{8}\M^{MN}\p_M \M^{PQ}\p_N \M_{PQ}\ \\[1.0ex]
&-\tfrac{1}{2} \M^{MN}\p_N \M^{KL} \p_L \M_{KM}\ -\M^{MN}\p_M \p_N \phi  \,, \\[1.5ex]
L_{4} \ = \ & \  -\tfrac{1}{12} \, \M^{MN} W_{4MN}    
\ + \tfrac{1}{6}  \, \p^P \M^{KL} \p_L W_{2KP}
+ \tfrac{1}{4}  \p_P G^K_1 (W_2) \M_K{}^P 
\\[1.0ex]
& \,  - \tfrac{1}{24}  \p_P \p_Q \M^{KL}  \p_K\p_L W_{0}^{PQ} \\[1.0ex]
& \ - \tfrac{1}{4} \p_N G^M_1 (\M) \p_M G_1^N (W_0) 
- \tfrac{1}{8}  \p_P \p_Q G^K_1 (\M) \p_K W_0^{PQ} - \tfrac{1}{8}  \p_P \p_Q G^K_1 (W_0) \p_K \M^{PQ} ,
\\[1.5ex]
L_{6} \ = \ & \  -\tfrac{1}{12} \, \M^{MN} W_{6MN}  
\ + \tfrac{1}{6}  \, \p^P \M^{KL} \p_L W_{4KP}
+ \tfrac{1}{4}  \p_P G^K_1 (W_4) \M_K{}^P \\[1.0ex]
& \,  - \tfrac{1}{24}  \p_P \p_Q \M^{KL}  \p_K\p_L W_{(2)}^{PQ} \\[1.0ex]
& \ - \tfrac{1}{4} \p_N G^M_1 (\M) \p_M G_1^N (W_2) 
- \tfrac{1}{8}  \p_P \p_Q G^K_1 (\M) \p_K W_{2}^{PQ} - \tfrac{1}{8}  \p_P \p_Q G^K_1 (W_{2}) \p_K \M^{PQ} \,.
\end{split} 
\ee
The results for the zero and two derivative part of the Lagrangian were given 
 in~\cite{Hohm:2013jaa,Hohm:2014xsa} and cannot be simplified further.    
One can quickly show that the last line of $L_{4}$ and $L_{6}$ vanish if we have zero
dilaton derivatives.   Also the last two terms in the first lines of $L_{4}$ and $L_{6}$
admit simplification.  Still keeping all terms,  we can simplify $L_{4}$ and $L_{6}$ to read
\be
\begin{split}
L_{4} \ = \ & \  -\tfrac{1}{12} \, \M^{MN} W_{4MN}    
\ + \tfrac{1}{12}  \, \p^M G_1^N (\M) W_{2MN}  
+ \tfrac{1}{6}  \M^{MK} W_{2K}{}^{N}  \partial_{MN}  \phi  \,  
\\[1.0ex]
& \,  - \tfrac{1}{24}  \p_P \p_Q \M^{KL}  \p_K\p_L W_{0}^{PQ} \\[1.0ex]
& \ + \tfrac{1}{4}  G^M_1 (\M) G_1^N (W_0) \partial_{MN}  \phi
+ \tfrac{1}{8} \bigl( G^K_1 (\M) W_0^{PQ} + G^K_1 (W_0) \M^{PQ} \bigr) \partial_{KPQ}\,  \phi   , 
\\[1.5ex]
L_{6} \ = \ & \  -\tfrac{1}{12} \, \M^{MN} W_{6MN}  
\ + \tfrac{1}{12}  \, \p^M G_1^N (\M) W_{4MN}  
+ \tfrac{1}{6}  \M^{MK} W_{4K}{}^{N}  \partial_{MN}  \phi \\[1.0ex]
& \,  - \tfrac{1}{24}  \p_P \p_Q \M^{KL}  \p_K\p_L W_{2}^{PQ} \\[1.0ex]
& \ + \tfrac{1}{4}  G^M_1 (\M) G_1^N (W_2) \partial_{MN}  \phi
+ \tfrac{1}{8} \bigl( G^K_1 (\M) W_2^{PQ} + G^K_1 (W_2) \M^{PQ} \bigr) \partial_{KPQ} \phi \,.
\end{split} 
\ee
The fourth and sixth derivative part of the Lagrangian, written explicitly in terms of $\M$ and $\phi$
are rather long.  
Since we will focus in this paper on gravity and $b$ field three-point amplitudes, we will ignore the dilaton. With dilaton fields set to zero a computation gives:
\be
\label{eq:L_6_phi0_cubic}
\begin{split}
L_{4}|_{\phi=0}\ =\ &\ \ 
\M^{MN}\ \bigl(\,    
\  \tfrac{1}{6} \p_{ML} \M^{PQ} \p_{PQ} \M_{N}{}^L\ 
-\tfrac{1}{12}  \p_{NP} \M^{LQ}\p_{ML} \M^P{}_Q \ 
\\[1.0ex]
& \hskip40pt + \tfrac{1}{12} \p_{MN} \M_{KQ} \p_{P}^K \M^{PQ} \ -\tfrac{1}{12} \p_{MP} \M^{PQ} \p_{NK} \M_{Q}{}^K 
\\[1.0ex]
& \hskip40pt +\tfrac{1}{3} \p_P \M_{M}{}^K \p_{NKQ} \M^{PQ} 
\ -\tfrac{1}{6} \p_M \M^{PQ} \p_{PK[N} \M_{Q]}{}^K
\ \bigr)
\\[1.0ex]
& \, +\p^{MP}\M_P{}^{N}\bigl( \tfrac{1}{6}\p_N \M^{KL} \p_L \M_{MK}\ -\tfrac{1}{6}\p_Q \M_{M}{}^K \p_K \M_{N}{}^Q\ + \tfrac{1}{12}\p_L \M^{KL} \p_K \M_{MN} \bigr), \\[2.0ex]
L_{6}|_{\phi=0}\ =\ &\ \M^{MN}\bigl(\  \tfrac{1}{48} \p_{MPQ}\M^{KL}\p_{NKL}\M^{PQ}\ 
+\tfrac{1}{24}\p_{MPQL}\M^{KL}\p_{NK}\M^{PQ}\  \\[1.0ex]
 & \hskip30pt-\tfrac{1}{24}\p_{PQKL}\M^{KL}\p_{MN}\M^{PQ} 
 \  +\tfrac{1}{12}\p_{MPKL}\M^{KL}\p_{NQ}\M^{PQ}\bigr)\\[1.0ex] 
&\, -\tfrac{1}{24}\p_{MNKL}\M^{KL}\bigl(\p_P\M^{PQ}\p_{Q}\M^{MN}\ -2\p_P\M^{MQ}\p_Q\M^{NP}\bigr)\\[1.0ex]
&\, -\tfrac{1}{24}\p_{NL}\M^{ML}\bigl( 2\p_{MK}\M^{PQ}\p_{PQ}\M_{N}{}^{K} +2 \p_{MPQ}\M^{KQ}\p_{K}\M^{P}{}_{N}\,    +    \ \p_{PQR}\M^{NR}\p_M\M^{PQ}\bigr).
\end{split}
\ee

\sectiono{Perturbative expansion of HSZ theory}
In this section we discuss the perturbative expansion of the Lagrangian obtained in the previous section around a {\em constant} background $\langle \M \rangle$ that  can be identified with 
a constant 
 generalized metric, as discussed in \cite{Hohm:2014xsa}. 
We define projected $O(D,D)$ indices as follows:
\be 
V_{\nin{M}}\ =\ P_M{}^{N}V_N,\ \ V_{\bar{M}}\ =\ \bar{P}_M{}^{N}V_N,
\ee
where the projectors are defined as:
\be 
P_{M}{}^N\ =\tfrac{1}{2}(\eta-\bar{\H})_{M}{}^N,\ \ \ \ \bar{P}_M{}^N\ 
=\tfrac{1}{2}(\eta+\bar{\H})_{M}{}^{N}  \,.
\ee
Here $\bar{\H}$ is the background, constant,  generalized metric. 
We expand the double metric $\M$ as follows:
\be 
\M_{MN}\ =\bar{\H}_{MN}+m_{MN}\ =\ \bar{\H}_{MN}\ +m_{\nin{M}\nin{N}}\ +m_{\nin{M}\bar{N}}\ +m_{\bar{M}\nin{N}}\ + m_{\bar{M}\bar{N}},\label{eq:Mexpansion1}
\ee
where we have decomposed the fluctuations $m_{MN}$ into projected indices. It was shown in \cite{Hohm:2014xsa} that the projections $m_{\bar{M}\bar{N}}$ and $m_{\nin{M}\nin{N}}$ are auxiliary fields and the physical part of the metric and the $b$-field fluctuations are encoded in $m_{\nin{M}\bar{N}}=m_{\bar{N}\nin{M}}$. To obtain the Lagrangian in terms of physical fields, we need to expand it in fluctuations and then eliminate the auxiliary fields using their equations of motion. To illustrate this procedure more clearly, and for ease of readability
we will write   
\be
a_{\nin{M}\nin{N}} \ \equiv \  m_{\nin{M}\nin{N}}\,,  \qquad    
a_{\bar{M}\bar{N}}  \ \equiv \  m_{\bar{M}\bar{N}}\,,
\ee
where the label $a$ for the field reminds us that it is auxiliary.  With this notation
 the $\M$ field expansion reads
\be 
\M_{MN} \  = \bar{\H}_{MN}\ +a_{\nin{M}\nin{N}}\ +m_{\nin{M}\bar{N}}\ +m_{\bar{M}\nin{N}}\ + a_{\bar{M}\bar{N}}\, .
\label{eq:Mexpansion}
\ee
Let us now carry out the procedure of elimination of auxiliary field
 explicitly for the two derivative part of the Lagrangian.

\subsection{Perturbative expansion of the two-derivative Lagrangian}

Let us use $L^{(i,j)}$ to denote the part of the Lagrangian with $i$ fields and $j$ derivatives. In what follows, we are only interested in the Lagrangian up to cubic order in fields, so we will ignore all terms with more than three fields. Also note that the Langrangian appears in the action multiplied with a factor of $e^{\phi}$. 
Using the expansion (\ref{eq:Mexpansion}) 
we see that the zero derivative Lagrangian $L_0$ has terms quadratic and cubic in field fluctuations:
\be
e^{\phi} L_{0}\ = \  \ L^{(2,0)}\ + L^{(3,0)}  + \cdots , 
\ee
where the dots denote terms quartic in fields and   
\be
\label{l20l30}
\begin{split} 
L^{(2,0)}\ = \ & \  \tfrac{1}{2}\,  a^{\,\nin{M}\nin{N}} a_{\nin{M}\nin{N}}\ -\tfrac{1}{2}\,  a^{\bar{M}\bar{N}}a_{\bar{M}\bar{N}},\\[1.0ex]
L^{(3,0)}\ =&\ -\tfrac{1}{2}\, a^{\, \nin{M}\nin{N}} m_{\nin{M}}{}^{\bar{P}} m_{\nin{N}\bar{P}}\ 
-\tfrac{1}{6} \, a^{\,\nin{M}\nin{N}} a_{\nin{M}}{}^{\nin{P}} a_{\nin{N}\nin{P}} \ 
-\tfrac{1}{2} \, a^{\bar{M}\bar{N}}m^{\nin{P}}{}_{\bar{M}} m_{\nin{P}\bar{N}}\ 
-\tfrac{1}{6} \, a^{\bar{M}\bar{N}} a_{\bar{M}
%un corrected index
}{}^{\bar{P}} a_{\bar{N}\bar{P}} \\[1.0ex]
&\ +\tfrac{1}{2}\phi \, \bigl (a^{\nin{M}\nin{N}} a_{\nin{M}\nin{N}}\
 - a^{\bar{M}\bar{N}}a_{\bar{M}\bar{N}} \bigr).
\end{split}
\ee
If we denote generically by $a$ an auxiliary field ($a_{\nin{M}\nin{N}}$ or $a_{\bar{M}\bar{N}}$) and
by $m$ the physical field $m_{\nin{M}\bar{N}}$,     
the structure of terms with auxiliary field that we find here is 
\be
\label{afielsm}
a^2 + a m^2 + a^3 + a^2 m  \,.     
\ee   
If we solve for the auxiliary field based on the above, to leading order we will find $a \sim m^2$. 
 The perturbative expansion for the two-derivative 
Lagrangian $L_2$ in (\ref{l0l2l4l6})   
 is  more involved. It decomposes into a quadratic and a cubic part in fluctuations:
\be 
e^{\phi}L_{2}\ =\ L^{(2,2)}\ +\ L^{(3,2)} + \cdots .
\ee
and we find 

      \be \label{eq:L2232}
	\begin{split}
	L^{(2,2)}\ =    & \ \tfrac{1}{2}\pd^{\bar{M}}m^{\nin{P}\bar{Q}}\pd_{\bar{M}}m_{\nin{P}\bar{Q}}
			+\tfrac{1}{2}\pd^{\nin{M}}m^{\nin{P}\bar{Q}}  \pd_{\nin{P}}m_{\nin{M}\bar{Q}}  
			-\tfrac{1}{2}\pd^{\bar{M}}m^{\nin{P}\bar{Q}}\pd_{\bar{Q}}m_{\nin{P}\bar{M}} \\[1.0ex]
			& -2m^{\nin{M}\bar{N}}\pd_{\nin{M}}\pd_{\bar{N}}\phi
			-2\phi\, \pd^{\bar{M}}\pd_{\bar{M}}\phi \  \\[1.0ex] 
			& +\tfrac{1}{4}\p^{\bar{M}} a^{\bar{P}\bar{Q}}\,  \p_{\bar{M}}a_{\bar{P}\bar{Q}}\ 
			+\ \tfrac{1}{4} \p^{\bar{M}} a^{\nin{P}\nin{Q}} \, \p_{\bar{M}}a_{\nin{P}\nin{Q}} \\[1,0ex]
			& +\tfrac{1}{2} \p^{\nin{M}} a^{\nin{P} \nin{Q}} \,\p_{\nin{Q}} a_{\nin{P}\nin{M}}\ 
			- \tfrac{1}{2} \p^{\bar{M}} a^{\bar{P} \bar{Q}} \, \p_{\bar{Q}} a_{\bar{P}\bar{M}} \,, \\[2.0ex]
	L^{(3,2)} \ =   & \ \ \ 
  			\tfrac{1}{2}m^{\nin{M}\bar{N}}
			\bigl( \pd_{\nin{M}}m^{\nin{P}\bar{Q}}\pd_{\bar{N}}m_{\nin{P}\bar{Q}}-\pd_{\nin{M}}m^{\nin{P}\bar{Q}}
			\pd_{\bar{Q}}m_{\nin{P}\bar{N}}-\partial_{\bar{N}} m^{\nin{P}\bar{Q}}\partial_{\nin{P}}m_{\nin{M}\bar{Q}}
			\bigr) \\[1.0ex]
			&\ \ \ +\tfrac{1}{2} \phi \bigl( \p_{\bar{M}} m^{\nin{P}\bar{Q}} \p^{\bar{M}} m_{\nin{P}\bar{Q}}\ -\p_{\bar{M}}
			m^{\nin{P}\bar{Q}} \p_{\bar{Q}} m_{\nin{P}}{}^{\bar{M}}\ +\p^{\nin{M}}m^{\nin{P}\bar{Q}} \p_{\nin{P}} m_{\nin{M}\bar{Q}})
			\\[1.0ex]
			& \  -\tfrac{1}{2} \bigl(  m_{\nin{M}}{}^{\bar{P}} m_{\nin{N}\bar{P}}\p^{\nin{M}}\p^{\nin{N}}\phi
			  - m^{\nin{P}}{}_{\bar{M}} m_{\nin{P}\bar{N}}\p^{\bar{M}}\p^{\bar{N}} \phi \bigr)
			\\[1.0ex]
			&\ \ \ -\phi^{2}\p_{\bar{M}}\p^{\bar{M}}\phi\ -2\phi m^{\nin{M}\bar{N}}\p_{\nin{M}}\p_{\bar{N}}\phi
			\ +\   L^{(3,2)}_{\text{aux}}\,  .
	   \end{split}
	\ee

where $L^{(3,2)}_{\text{aux}}$ denotes the terms that contain at least one auxiliary field.
The precise expression for these terms will not be needed. 
Note, however, from  $L^{(2,2)}$ that we have terms of the form 
\be
\label{moreafiels}
 \p a \p a \,,    
 \ee 
and from $L^{(3,2)}$ terms that couple an $a$ field to two fields in a term with two derivatives. 

Next, we eliminate the auxiliary fields from  
the total Lagrangian with three or less fields and at most two derivatives.   
\be
\label{ltotal-sofar} 
L^{(\leq 3, 2) } \ = \ L^{(2,0)} + L^{(3,0)} + L^{(2,2)} + L^{(3,2)} \,, 
\ee
From the terms in (\ref{l20l30}) and (\ref{eq:L2232}), denoted schematically  
in (\ref{afielsm}) and (\ref{moreafiels}),  we now find:
\be
\label{auxfieldsolve}
a_{\bar{M}\bar{N}}\ = \  -\tfrac{1}{2} m^{\nin{P}}{}_{\bar{M}} m_{\nin{P}\bar{N}} \, + \cdots  \,,  \qquad 
a_{\nin{M}\nin{N}}\ =  \ \tfrac{1}{2} m_{\nin{M}}{}^{\bar{P}} m_{\nin{N}\bar{P}} \,  + \cdots \, .
\ee
where the dots denote terms with at least two fields and at least two derivatives.    
Now, we plug this solution for the auxiliary field into the Lagrangian $L^{(\leq 3, 2)}$  
and keep only terms with two derivatives and up to cubic order in physical fields.  The terms indicated by dots in (\ref{auxfieldsolve})
do not contribute; they always lead to terms with at least four fields or at least four
derivatives. 
Nor does $L^{(3,2)}_{\text{aux}}$ lead to any contributions.  In fact, 
 most of the terms involving auxiliary fields do not contribute.
 After a short  
 computation, we obtain the following two derivative Lagrangian completely in terms of the physical fields:
\be
		\begin{split}
		L^{(\leq 3, 2)}   
				\ =\ & \ \ \ \tfrac{1}{2}\pd^{\bar{M}}m^{\nin{P}\bar{Q}}\pd_{\bar{M}}m_{\nin{P}\bar{Q}}
				+\tfrac{1}{2}\pd^{\nin{M}}m^{\nin{P}\bar{Q}}\pd_{\nin{P}}m_{\nin{M}\bar{Q}}
				-\tfrac{1}{2}\pd^{\bar{M}}m^{\nin{P}\bar{Q}}\pd_{\bar{Q}}m_{\nin{P}\bar{M}} \ \\[1.0ex]
				& -2m^{\nin{M}\bar{N}}\pd_{\nin{M}}\pd_{\bar{N}}\phi-2\phi\pd^{\bar{M}}\pd_{\bar{M}}\phi \\[1.0ex]
				& +\tfrac{1}{2}m^{\nin{M}\bar{N}}\bigl( \pd_{\nin{M}}m^{\nin{P}\bar{Q}}\pd_{\bar{N}}m_{\nin{P}\bar{Q}}
				-\pd_{\nin{M}}m^{\nin{P}\bar{Q}}\pd_{\bar{Q}}m_{\nin{P}\bar{N}}
				-\partial_{\bar{N}} m^{\nin{P}\bar{Q}}\partial_{\nin{P}}m_{\nin{M}\bar{Q}} \bigr) \\[1.0ex]
				& +\tfrac{1}{2} \phi \bigl( \p_{\bar{M}} m^{\nin{P}\bar{Q}} \p^{\bar{M}} m_{\nin{P}\bar{Q}}\
				 -\p_{\bar{M}}m^{\nin{P}\bar{Q}} \p_{\bar{Q}} m_{\nin{P}}{}^{\bar{M}}\
				 +\p^{\nin{M}}m^{\nin{P}\bar{Q}} \p_{\nin{P}} m_{\nin{M}\bar{Q}})\\[1.0ex]
				& -\tfrac{1}{2} \bigl(  m_{\nin{M}}{}^{\bar{P}} m_{\nin{N}\bar{P}}\p^{\nin{M}}\p^{\nin{N}}\phi  
				- m^{\nin{P}}{}_{\bar{M}} m_{\nin{P}\bar{N}}\p^{\bar{M}}\p^{\bar{N}} \phi \bigr)
				\\[1.0ex]
				& -\phi^{2}\p_{\bar{M}}\p^{\bar{M}}\phi\ 
				-2\phi m^{\nin{M}\bar{N}}\p_{\nin{M}}\p_{\bar{N}}\phi \ .
				\end{split}\label{eq:Lcubicm}
		\ee

Next, we write the action in terms of double field theory (or string field theory)  
 variables $e_{ij}$. The way to translate from  $m_{\nin{M}\bar{N}}$ variables to   $e_{ij}$ variables is explained in Sec.~5.3 of \cite{Hohm:2014xsa}.
  Here is the rule that follows:  Convert all barred and under-barred indices into latin indices respecting
 the contractions,  replacing $m$ by $e$, underbar derivatives by $D$ and barred derivatives by $\bar D$, and multiply by a coefficient that is the product of a factor of 2 for each $m$ field,
a factor of $+\tfrac{1}{2}$ for 
 each barred contraction, and  a factor of $-\tfrac{1}{2}$ for each under-barred contraction.
As an example,
consider the second term on the first line of (\ref{eq:Lcubicm}), after integration by parts, it becomes:
\be 
\tfrac{1}{2}\pd^{\nin{M}}m^{\nin{P}\bar{Q}}\pd_{\nin{P}}m_{\nin{M}\bar{Q}}\ = \tfrac{1}{2}\pd_{\nin{P}}m^{\nin{P}\bar{Q}}\pd^{\nin{M}}m_{\nin{M}\bar{Q}}\ \quad \to \  
\tfrac{1}{2} \cdot 2^2 \cdot \ \tfrac{1}{2} \bigl(-\tfrac{1}{2}\bigr)^2
D_p e^{pq} D^m e_{mq} 
\ = \tfrac{1}{4} D_p e^{pq} D^m e_{mq}.   
\ee
Using this technique for all the terms appearing in the Lagrangian (\ref{eq:Lcubicm}) we obtain:
\begin{equation}\label{eq:L(2,2)CSFT1}
\begin{split}
L^{(\leq 3, 2)}  
\ =\ 
&\ \ \ \tfrac{1}{4}\bkt{e^{ij}\bar{D}^2e_{ij}+\bkt{D^ie_{ij}}^2+\bkt{\bar{D}^ie_{ij}}^2}+e^{ij}D_i\bar{D}_{j}\phi-\phi\bar{D}^2\phi.
\\ 
\ &+\tfrac{1}{4}e_{ij}
\Bigl(     
D^{i}e_{kl} \bar{D}^je^{kl}
-D^ie_{kl} \bar{D}^le^{kj}
-D^ke^{il} \bar{D}^je_{kl}     
\Bigr)   
 \\[1.0ex] 
&-\tfrac{1}{4} \phi \left( 
\bigl(D^i e_{ij}\bigr)^2 +\bigl(\bar{D}^{j} e_{ij}\bigr)^2  
+\tfrac{1}{2}\bigl(D_{k} e_{ij}\bigr)^2  
+\tfrac{1}{2} \bigl(\bar{D}_k e_{ij}\bigr)^2 
+2 e^{ij} \bigl(D_iD^ke_{kj}\ + \bar{D}_j\bar{D}^k e_{ik}\bigr)\right)\\[1.0ex]
&\ +\phi e_{ij}D^i\bar{D}^j\phi\ -\tfrac{1}{2} \phi^2 \bar{D}^2 \phi.
\end{split}
\end{equation}
With the identification $\phi=-2d$   
the above cubic Lagrangian becomes precisely the double field theory Lagrangian 
 in equation (3.25) of \cite{Hull:2009mi}. 
From the quadratic part of the above action, we see that the kinetic term of $\phi$ has wrong sign. This is, because the action (\ref{eq:L(2,2)CSFT1}) is in the string frame 
and $\phi$ is not the physical dilaton. 
To obtain the action in terms of physical fields $\hat e_{ij}$ and $\hat \phi$ that
decouple at the quadratic level, 
 we need a field re-definition. 
 Physical fields $\hat e_{ij}$ and $\hat \phi$ are obtained in the Einstein frame as a linear combination of $e_{ij}$ and $\phi$. We write schematically:   
\be 
 e_{ij}\ \sim \hat e_{ij}\ +\hat\phi \, \eta_{ij}, \qquad   
\phi \sim \hat\phi \ + \hat e_i^{\ i}.
\ee
If we are looking for pure gravitational three-point amplitudes the first redefinition need not be performed in the action, as it  would give rise to terms that involve the dilaton.  The second one
is not needed either, since on-shell gravitons have traceless polarizations.

After solving the strong constraint by setting $\tilde{\p}^i\ =\ 0$ and setting the dilaton to zero, the above Lagrangian becomes:
\begin{equation}
\label{eq:L(2,2)CSFT}
L^{(\leq 3, 2)}\Big|_{\phi=0}\ =\    
 \tfrac{1}{4}\bkt{e^{ij}\p^2e_{ij}+2\bkt{\p^ie_{ij}}^2} + \ \tfrac{1}{4}e_{ij}\bkt{\p^{i}e_{kl}\p^je^{kl}-\p^ie_{kl}\p^le^{kj}-\p^ke^{il}\p^je_{kl}}. 
\end{equation}
For an off-shell three-point vertex all terms in the cubic Lagrangian must be kept.
But for the computation of {\em on-shell}  three-point amplitudes we may 
use the on-shell conditions to simplify the cubic Lagrangian. 
These  conditions can be stated as follows in terms of $e_{ij}$.
\be 
\p^{i}e_{ij}\ =\p^je_{ij}= 0\, , \ \ \ \ \ \ \p_{i} e^{\cdot \cdot} \p^i e^{\cdot \cdot} \cdots\ =0.\label{eq:onshelle}
\ee
The first condition
is transversality and the second condition follows from the momentum conservation and masslessness. 
For the cubic terms in (\ref{eq:L(2,2)CSFT}) the on-shell conditions do not lead to any further 
simplification and we record: 
\be \label{eq:LtwoDerOnShell}
\begin{split}
L^{(3,2)}\Big|_{\phi=0,\text{ on-shell}}= \tfrac{1}{4}e_{ij}\bkt{\p^{i}e_{kl}\p^je^{kl}-\p^ie_{kl}\p^le^{kj}-\p^pe^{iq}\p^je_{pq}}. 
\end{split}
\ee
Three-point on-shell amplitudes can now be computed from this expression.

\subsection{General treatment of auxiliary fields}  
 
Here we argue that for the purposes of three-point on-shell amplitudes and, with the dilaton set to zero, the auxiliary field does not affect the Lagrangian and can safely be ignored.
To prove the claim we must use on-shell conditions (\ref{eq:onshelle}):  we will argue that
any contribution from auxiliary fields vanishes upon use of these conditions.
 It is straightforward to translate these on-shell conditions in terms of the double metric fluctuations. They can be written as:
\be 
\p_{\nin{M}}m^{\nin{M}\bar{N}}\ =\ \p_{\bar{N}}m^{\nin{M}\bar{N}}\ =\ \p_{\bar{M}}m^{\cdot\cdot}\p^{\bar{M}}m^{\cdot\cdot}\cdots\ =\ 0\, .\label{eq:Onshell_m}
\ee
Setting all dilatons to zero, the only physical field is $m_{\nin{M}\bar{N}}$, which we symbolically
represent by $m$.  The most general form of the Lagrangian involving at least one auxiliary field 
is as follows:
\be 
L[a,m]\ =\  a  m \ +a^2 \ + a^3\ +a^2m\ +am^2 \,.   
\label{eq:AuxSchematic3}  
\ee
Since the theory is cubic in $\M$ and the dilaton is set to zero, this is all there is.
In here we are leaving derivatives implicit; all the above terms 
can carry up-to six derivatives.  As we have seen before, there is no $a m$ coupling with zero derivatives
nor with two derivatives.  
Let us now see that no such term exists that does not
vanish using the on-shell conditions.  The general term of this kind would
be
\be  
m_{\bar{M}\bar{N}} \bigl( \cdots    m_{\nin{P} \bar{Q}} ) \,, 
\ee
where the dots represent derivatives or metrics $\eta$ that contract same type indices, barred or un-barred.  These are required to contract all indices and yield an $O(D,D)$ invariant.  Since integration
by parts is allowed we have assumed, without loss of generality that all derivatives are acting on the
physical field.  Since the un-barred index $P$ is the only un-barred index, it must be contracted
with a derivative.  Thus the term must be of the form 
\be  
m_{\bar{M}\bar{N}} \bigl( \cdots   \p^{\nin{P}} m_{\nin{P} \bar{Q}} ) \,.
\ee
Regardless of what we do to deal with the other barred indices, we already see that this coupling vanishes
using the on-shell conditions, proving the claim.

 The Lagrangian (\ref{eq:AuxSchematic3}) then reduces to the following:
 \be 
 L[a,m]\ =\ a^2 \ + a^3\ +a^2m\ +am^2\, .  \label{eq:AuxSchematic} 
 \ee 
 
 The equation of motion for the auxiliary field is, schematically,  $a \sim  m^2  + am +  a^2$, which
 implies that a perturbative solution in powers of physical fields begins with terms quadratic on the
 physical fields.  Thus we write
 \be 
a (m)\ =\  a_{2}(m) \ + a_{3} (m) + \cdots \,,   \label{eq:AuxSol}   
\ee
where dots indicate terms with quartic or higher powers of $m$.  But now it is clear that substitution
back into (\ref{eq:AuxSchematic}) can only lead to terms with quartic or higher powers of $m$.
This concludes our argument that the elimination of auxiliary fields is not required
for the computation of on-shell three-point amplitudes for metric and $b$ fields.

\subsection{Higher-derivative Lagrangian and on-shell amplitudes}\label{higderlag}

In this subsection we perform the perturbative expansion of the four and six derivative Lagrangian
and compute the on-shell three-point amplitudes. 
We use the on-shell conditions (\ref{eq:Onshell_m}) and ignore the auxiliary field in light of our earlier discussion. We note that 
\be
\label{eq:onshellDouble} 
\begin{split}
\p^{M}\M_{M \bar N}\ =\   &\  \p^{\bar{M}} a_{\bar{M}\bar{N}}\  +\ \p^{\nin{M}}m_{\nin{M}\bar{N}}\,, \\[1.0ex]
\p^{M}\M_{M\nin{N}}\ = \  &\  \p^{\bar{M}}m_{\bar{M}\nin{N}}\ + \p^{\nin{M}}a_{\nin{M}\nin{N}}\ \,. 
\end{split}
\ee

Since we are allowed to set auxiliary fields to zero and to use the on-shell conditions
(\ref{eq:Onshell_m}), both 
$ \p^{M}\M_{M \bar N}$ and $\p^{M}\M_{M\nin{N}}$ can be set to zero, and as a result, we are
allowed to set
\be
\label{eq:onshellM}
\p^{M}\M_{M  N} \ \to \ 0 \,,
\ee
in simplifying the higher-derivative cubic interactions!  This is a great simplification.

Now we use (\ref{eq:onshellM}) in the four derivative Lagrangian  $L_4$ given
in (\ref{eq:L_6_phi0_cubic}).  
Only the terms on the first line survive and we get:  
\be 
\ L_{4}\Big|_{\phi=0}\ =\    
\tfrac{1}{6}\M^{MN}\p_{NL}\M^{PQ}\p_{PQ}\M_M{}^{L}\   
-\tfrac{1}{12}\M^{MN}\p_{NP}\M^{LQ}\p_{ML}\M^{P}{}_Q\,   . \label{eq:L4Pi0on_shell}
\ee
Now we plug in the expansion (\ref{eq:Mexpansion}) and keep only the cubic terms which do not vanish on-shell. After a short computation we obtain the four derivative cubic Lagrangian in terms of the physical fields 
\be 
 L^{(3,4)}\Big|_{  {\phi=0}\atop{\text{on-shell}} } =\   \tfrac{1}{3}\ 
 m^{\nin{M}\bar{N}} 
 \Bigl(
 \p_{\bar{N}\bar{L}} m^{\nin{P}\bar{Q}}\ \bigl[ 
 \p_{\nin{P}\bar{Q}}m_{\nin{M}}{}^{\bar{L}}\ 
 -\tfrac{1}{2} \p_{\nin{M}\bar{Q}}m_{\nin{P}}{}^{\bar{L}} \bigr] 
 +\ \p_{\nin{M}\nin{L}} m^{\nin{P}\bar{Q}} 
 \bigl[ \p_{\nin{P}\bar{Q}} m^{\nin{L}}{}_{\bar{ N}}  -\tfrac{1}{2}
  \p_{\bar{N}\nin{P}} 
  m^{\nin{L}}{}_{\bar{Q}} \bigr] \   
\Big) \label{eq:L4physical}.
\ee
Translating this to $e$ fluctuations (three $m$'s and 5 contractions):
\be 
\label{eq:L4physical9}
 \begin{split}
 L^{(3,4)}\Big|_{  {\phi=0}\atop{\text{on-shell}}  } =&\   \tfrac{1}{12}\ 
 e^{ij} 
 \Bigl(
 \p_{jl} e^{pq}\ \bigl[ 
 \p_{pq}e_{i}{}^{l}\ 
 -\tfrac{1}{2} \p_{iq} e_{p}{}^{l} \bigr] 
 -\ \p_{il} e^{pq} 
 \bigl[ \p_{pq} e^{l}{}_{j}  -\tfrac{1}{2}
  \p_{pj} e^l{}_q \bigr] \      
\Big) \,. 
 \end{split}
\ee
Using integration by parts and the gauge conditions this simplifies into:   
\be 
\label{4der-3pt}
L^{(3,4)}\Big|_{  {\phi=0}\atop{\text{on-shell}}  } \ = \ \tfrac{1}{8} \ e_{ij}\bigl(
\p^{jq} e_{kl} \p^{kl} e^i{}_q     
-  \p^{ip} e_{kl} \p^{kl} e_p{}^j 
\bigr) \,,
\ee
and written  in terms of the metric and $b$ field fluctuations using $e_{ij} =\ h_{ij}+b_{ij}$:
\be  
L^{(3,4)}\Big|_{  {\phi=0}\atop{\text{on-shell}}  } \ =\ \tfrac{1}{2}\,  
b^{ij} \, \p_{jl} h^{mn}\, \p_{mn} h^l{}_i   \,.
\ee
A short computation confirms that this result
 is precisely produced by the on-shell perturbative evaluation of 
 the action 
 \be
 \label{l34hz}
 { L}^{(3,4)} = -\tfrac{1}{2} H^{ijk} \Gamma^q_{ip} \p_j \Gamma ^p_{kq}\,, 
 \ee  
given in equation (3.23) of  \cite{Hohm:2015mka} and arising from the expansion of the 
kinetic term for the Chern-Simons improved field strength $\widehat H$.   
 There is no Riemann-squared term appearing, as has been argued before.

In the six-derivative Lagrangian $L_6$ given in (\ref{eq:L_6_phi0_cubic})   
only the first term survives after we impose the on-shell condition.
Integrating by parts the $\p_N$ derivative we have 
\be 
 L_{6}\Big|_{  {\phi=0}\atop{\text{on-shell}}  } \ =  
\ -  \tfrac{1}{48}\, \M^{MN} \p_{MNPQ} \M^{KL} \p_{KL} \M^{PQ}.  
\ee
Using the $\M$ field expansion and keeping only cubic terms which are non-vanishing on-shell, we get:
\be   
 L^{(3,6)}\Big|_{  {\phi=0}\atop{\text{on-shell}}  } \ =\ -\tfrac{1}{6}\,  m^{\nin{M}\bar{N}} 
 \p_{\nin{M}\bar{N}\nin{P}\bar{Q}}
 m^{\nin{K}\bar{L}}
 \p_{\nin{K}\bar{L}} m^{\nin{P}\bar{Q}}.
\ee
In term of $e_{ij}$ this takes the form:
\be
\label{6der-3pt} 
 L^{(3,6)}\Big|_{  {\phi=0}\atop{\text{on-shell}}  } \ =\tfrac{1}{48} \, e_{ij} \, \p^{ijpq} e_{kl} \, \p^{kl} e_{pq}\, .
\ee 

The structure of the six-derivative term is such that only the symmetric part of $e_{ij}$ contributes. In terms of the metric fluctuations we get:
\be 
 L^{(3,6)}\Big|_{  {\phi=0}\atop{\text{on-shell}}  } \ =\tfrac{1}{48} \, h_{ij} \, \p^{ijpq} h_{kl} \, \p^{kl} h_{pq}\, .\label{eq:sixterm}
\ee  
This term is produced by the perturbative on-shell evaluation of the following 
Riemann-cubed term:  
\be 
\label{riemanncubed}
-\tfrac{1}{48} R_{ij}^{\ \ kl}R_{kl}^{\ \ pq}R_{pq}^{\ \ ij},
\ee
where the linearized Riemann tensor is:  
$ % \be 
R_{ijkl} = \tfrac{1}{2}\left( \p_{jk} h_{il}\ +\p_{il} h_{jk}\ - \p_{ki} h_{jl}\ - \p_{jl} h_{ik}\right).
$  %\ee
A short computation then shows:
\be 
-\tfrac{1}{48} R_{ij}^{\ \ kl}R_{kl}^{\ \ pq}R_{pq}^{\ \ ij} \Big|_{\text{on-shell}}\ = -\tfrac{1}{48} \p^{lq}h_{ij}\p^{pj} h_{kl} \p^{ki} h_{pq}\, ,
\ee
which gives precisely the term (\ref{eq:sixterm}) after integration by parts.

Collecting our results (\ref{eq:LtwoDerOnShell}), (\ref{4der-3pt}) and (\ref{6der-3pt}) for the cubic 
interactions with two, four, and six derivatives, we have:
\be \label{total3pt}
\begin{split}
L_3\Big|_{  {\phi=0}\atop{\text{on-shell}}  } \ =  & \  \  \ \tfrac{1}{4}e_{ij}
\, \Bigl[  \ \ \p^{i}e_{kl}\p^je^{kl}-\p^ie_{kl}\p^le^{kj}-\p^pe^{iq}\p^je_{pq}   \\ 
&  \hskip30pt  +  \tfrac{1}{2} \alpha' \  \bigl(    
\p^{jq} e_{kl} \p^{kl} e^i{}_q   
 -  \p^{ip} e_{kl} \p^{kl} e_p{}^j 
\ \bigr)  
+ \tfrac{1}{12} \, \alpha'^2   
 \, \p^{ijpq} e_{kl} \, \p^{kl} e_{pq} \ \Bigr] \, ,
\end{split}
\ee
where we have made explicit the $\alpha'$ factors in the various contributions.  
To compute the on-shell amplitude we pass to momentum space.  We need not concern
ourselves with overall normalization; all that matters here is the relative numerical factors between
the two, four, and six-derivative terms. 
 We thus have an on-shell amplitude ${\cal A}$ 
proportional to
\be
\begin{split}
 {\cal A} \ =  \ e_{1ii'} e_{2jj'} e_{3kk'}  
\, \Bigl[ &  \ \ -k_2^i k_3^{i'} \, \eta^{jk} \eta^{j'k'}  +k_2^i k_3^{j'} \, \eta^{jk} \eta^{i'k'}\ 
+k_2^k k_3^{i'} \, \eta^{ij} \eta^{j'k'} \   + \hbox{permutations}   \\ 
&  \hskip7pt  +  \tfrac{1}{2} \alpha' \,  \bigl(    
k_2^{i'} k_2^{k'} k_3^j k_3^{j'}\, \eta^{ik} \  -\ k_2^{i} k_2^k k_3^j k_3^{j'}\, \eta^{i'k'}
 \bigr) \   + \hbox{permutations} \\[1.0ex] 
&  \hskip7pt - \tfrac{1}{12} \,\alpha'^2 \,  \, k_2^i k_2^{i'} k_2^k k_2^{k'}  \, k_3^j k_3^{j'}  \   + \hbox{permutations} \ \Bigr] \, ,    
\end{split} 
\ee
where we have used three different lines to list the terms with two, four, and six derivatives.
By `permutations' here we mean adding, in each line, the five copies with index permutations
required to achieve full Bose symmetry.
In order to show that the above has the conjectured factorized form we must rewrite the
momentum factors in terms of momentum differences $k_{12}, k_{23},$ and $k_{31}$.  
This is possible because momentum factors must contract
with polarization tensors, and using momentum conservation and transversality ensure they
can  be converted into momentum differences.  For example,
\be 
e_{2jj'} k_1^{j'}\ =\ \tfrac{1}{2}e_{2jj'} (k_{1}^{j'} + k_1^{j'})  \ = \tfrac{1}{2}e_{2jj'} (k_{1}^{j'} -k_2^{j'} -k_3^{j'} ) \ = \ 
 -\tfrac{1}{2}e_{2jj'} k_{31}^{j'}.
\ee
After rewriting all momenta as momentum differences the sum over permutations simplify
and with modest work one can show that the two, four, and six derivative terms 
can be written as sum of products of the $T$ and $W$ tensors introduced in (\ref{TW}).
Indeed, making use of (\ref{partial-amplitudes})  
 one finds, 
\be
\begin{split}
 {\cal A} \ =  \ & \tfrac{1}{2} e_{1ii'}(k_1) e_{2jj'}(k_2) e_{3kk'}(k_3) 
\, \Bigl[   \ \   T^{ijk}T^{i'j'k'} \  +  \ (W^{ijk}\, T^{i'j'k'}\ - T^{ijk}W^{i'j'k'}) \  - W^{ijk}\, W^{i'j'k'}
\ \Bigr] \\
\ =  \ & \tfrac{1}{2} e_{1ii'}(k_1) e_{2jj'}(k_2) e_{3kk'}(k_3) 
\,   ( T^{ijk} + W^{ijk})( T^{i'j'k'} - W^{i'j'k'}) \,, 
\end{split} 
\ee
in agreement with (\ref{eq:DFTprediction}) and thus proving the claimed factorization.   

\sectiono{Conclusions and remarks}

Our work has determined the form (\ref{CubicCov}) of the gauge invariant HSZ action 
  that reproduces the on-shell cubic amplitudes of the  theory.   
The ${\cal O}(\alpha')$ terms arise from the kinetic term for the  
three-form curvature $\widehat H$, which contains the Chern-Simons correction. 
Our work in section~\ref{higderlag} reconfirmed that the cubic on-shell four-derivative couplings arise correctly -- see (\ref{l34hz}).  
The kinetic term $\widehat H^2$ also contains  ${\cal O}(\alpha'^2)$ contributions, but those would only
affect six and higher-point amplitudes.   The full HSZ action may contain other  ${\cal O}(\alpha')$
terms that do not contribute to three-point  amplitudes.   
The action includes the 
Riemann-cubed term derived in (\ref{riemanncubed}).   
Its coefficient is 
minus the coefficient of the same term in bosonic string theory.   
In bosonic string theory there is also a non-zero `Gauss-Bonnet' Riemann-cubed term, 
but its presence can only
be seen from four-point amplitudes~\cite{Metsaev:1986yb}.  Neither the Riemann-cubed nor its
related Gauss-Bonnet term are present in heterotic string theory.  It would be interesting to see if 
the cubic-curvature Gauss-Bonnet interaction is present in HSZ theory.   
The physical effects of Riemann-cubed
interactions were considered in~\cite{Camanho:2014apa} and, 
regardless of the sign of
the term,  they lead to causality violations that require the existence of new particles. 

The action (\ref{CubicCov}), while exactly gauge invariant, is unlikely  to be
 {\em exactly} duality invariant.  It is not, after all, the full action for HSZ theory.  
Reference~\cite{Hohm:2015doa} showed  that the action (\ref{CubicCov}), without the
Riemann-cubed term, is not duality invariant to order $\alpha'$ squared.   
It may be possible to use the methods in~\cite{Hohm:2015doa}
to find out what other terms (that do not contribute to cubic amplitudes) are needed for duality invariance to order $\alpha'$ squared. We continue to expect that, 
in terms of a metric and a $b$-field,   
 an action with infinitely many terms is
required for exact duality invariance.

We have not attempted to compute dilaton amplitudes from HSZ theory.  There 
is no in-principle obstacle, and such computation could be done working in the Einstein frame.
The graviton and dilaton fluctuations $(\hat h_{\mu\nu}, \hat\phi)$ with standard, decoupled, kinetic terms are linear combinations of the fluctuations $(h_{\mu\nu}, \phi)$ that
we  use.  These redefinitions must be performed  to compute physical
dilaton amplitudes.  They were not needed to compute gravity and $b$-field amplitudes
because $h_{\mu\nu}$ differs from $\hat h_{\mu\nu}$  only by  dilaton dependent terms
and the dilatons differ from each other by traces of $h$, which do not contribute for
on-shell three-point amplitudes.   

The computation of quartic amplitudes in HSZ theory is clearly a very interesting challenge.
World-sheet methods may give an efficient way
to obtain answers. 
It is still important, however, to develop techniques to compute amplitudes in a 
theory with a double metric. 
The HSZ action  is not uniquely  
fixed by the gauge structure of the theory~\cite{Hohm:2013jaa}: one can add  
higher-order gauge-invariant products of the tensor field $\T$ which are
expected to modify quartic and higher-order amplitudes.
 In those theories, the spacetime action would be the natural tool to 
compute amplitudes, and one could wonder how the conformal field theory method
would work.  In this paper we have taken the first steps in the computation
of amplitudes  starting from a theory with a double metric. The computation of 
four-point amplitudes and of amplitudes that involve dilatons would be significant progress.

It is natural to ask to what
degree global duality determines the classical effective action for the massless fields of string theory.  Additionally, given an effective field theory of metric, $b$-field and dilaton, it is also natural to ask if the theory has a duality symmetry. HSZ theory is useful as it is the simplest gravitational theory with higher derivative corrections and exact global duality. By investigating
HSZ theory we will better understand the constraints of duality and its role in the effective 
field theory of strings. 

\section*{Acknowledgments} 
We thank Olaf Hohm for collaboration in early stages of this work  
and for useful comments. This work is supported by the U.S. Department of Energy under grant Contract Number  DE-SC0012567.

\end{document}